\documentclass[aps,pre,showpacs,amsfonts,amssymb,amsmath,floatfix,twocolumn,
floats,nobalancelastpage]{revtex4}

\usepackage{graphicx}

\begin{document}          

\title{Update statistics in conservative parallel discrete 
event simulations of asynchronous systems}

\author{A. Kolakowska}
\email{alicjak@bellsouth.net}
\author{M. A. Novotny}
\email{man40@ra.msstate.edu}
\affiliation{Department of Physics and Astronomy, and the ERC 
Center for Computational Sciences, 
P.O. Box 5167, Mississippi State, MS 39762-5167}
\author{Per Arne Rikvold}
\email{rikvold@csit.fsu.edu}
\affiliation{School of Computational Science and Information Technology, 
Center for Materials Research and Technology, and Department of Physics, 
Florida State University, Tallahassee, FL 32306-4120}

\date{\today}

\begin{abstract}                           
We model the performance of an ideal closed chain of $L$ 
processing elements that work in parallel in an asynchronous 
manner. Their state updates follow a generic conservative algorithm. 
The conservative update rule determines the growth of a virtual 
time surface. The physics of this growth is reflected in the 
utilization (the fraction of working processors) and in  
the interface width. We show that it is possible to make an 
explicit connection between the utilization and the microscopic 
structure of the virtual time interface. We exploit this connection 
to derive the theoretical probability distribution of updates in the system 
within an approximate model.  
It follows that the theoretical lower bound for the computational 
speed-up is $s=(L+1)/4$ for $L \ge 4$. Our approach uses simple 
statistics to count distinct surface configuration classes 
consistent with the model growth rule. It enables one to compute 
analytically microscopic properties of an interface, 
which are unavailable by continuum methods.
\end{abstract}


\maketitle

\section{INTRODUCTION \label{intro}}

In discrete event simulations a physical system with stochastic 
dynamics is modeled on a lattice of discrete points, and 
changes of its state are viewed as discrete events in time. 
Physical processes interact with each other at various 
points in simulation time. The stochastic nature of these 
interactions makes it difficult to utilize a parallel 
computing environment to the fullest extent because {\it a priori} 
there is no global clock to synchronize physical processes. 
Examples of such complex systems with underlying asynchronous dynamics 
come from a wide range of fields, such as, activated processes 
in chemistry, contact processes in epidemiology and ecology models, 
population dynamics, finance markets, 
communication networks and internet traffic, to mention a few. 
In physics an important example is an interacting spin system, 
where stochastic processes can be simulated 
with a dynamic Monte Carlo approach. Until recently a common belief in the physics 
community was that even the simplest random-site update Monte Carlo 
schemes \cite{BH97} were inherently serial. A popular parallelization 
technique  for these systems is the so-called ``trivial parallelization", 
in which each processor carries a copy of the full system. 
An obvious limitation of this technique is imposed by the 
memory requirement, which may exceed available resources for a 
large-scale simulation. In non-trivial parallelization, a system 
is spatially partitioned into subsystems, and each subsystem is 
placed on a different processor. In other words, in this way 
{\it physical processes} and physical interactions between subsystems 
are mapped to {\it logical processes} and logical dependences 
between processing elements. Each logical process manages the state 
of the assigned physical subsystem and progresses in its own local 
virtual time (LVT). The asynchronous nature of the physical dynamics 
implies an asynchronous system of logical processes where discrete 
events are not synchronized by a global clock. Logical processes 
execute concurrently and exchange time-stamped messages 
to perform state updates of the entire physical system being 
simulated. A sufficient condition for preserving causality in 
simulations (the so-called local causality constraint) requires that 
each logical process works out the received messages from other 
logical processes in nondecreasing time-stamp order 
\cite{CM79,Fuj90}.

Parallel discrete event simulations (PDES) are classified in 
two broad categories: conservative PDES and 
optimistic PDES. In conservative PDES, originally studied 
by Chandry and Misra \cite{CM79,Mis86} and introduced by Lubachevsky 
in the study of dynamic Ising spin systems \cite{Lub87,Lub88}, an 
algorithm does not allow a logical process to advance its LVT 
(i.e., to proceed with computations) until it is certain that no 
causality violation can occur. In the conservative update scenario 
a logical process may have to be blocked and wait to ensure 
that no message with a lower time-stamp is received later. 
Recent physics applications of conservative PDES in modeling 
magnetization switching \cite{KNR99}, ballistic particle 
deposition \cite{LP96}, and a dynamic phase transition in 
highly anisotropic thin-film ferromagnets \cite{KWRN01,KRN02}  
suggest that the conservative algorithm should be very 
efficient in simulating the dynamics of complex systems with 
short-range interactions. In optimistic PDES 
\cite{Jeff85,DR90,PS91,Stei93,FC94}, originated by Jefferson's 
time warp algorithm \cite{Jeff85}, an algorithm allows a logical 
process to advance its LVT regardless of the possibility of a causality 
error that may happen in the case of receiving a message with a lower 
time-stamp than the local clock. The optimistic scenario detects 
causality errors and provides a recovery procedure from 
the violation of the local causality constraint by rolling 
back the events that have been processed prematurely. Although 
there are no general performance studies to date that would provide an 
unbiased comparison of the two groups of algorithms, a common 
perception is that an optimistic PDES should outperform a conservative PDES. 
However, in the context of physics applications to Ising spin 
systems, recent numerical studies by Sloot {\it et al}. \cite{SOS01} 
demonstrate that near the Ising critical temperature, where 
long-range correlations occur in the physical spin system being 
modeled, the computational complexity of an optimistic PDES and 
the physical complexity of the modeled system are entangled, 
leading to a nonlinear increase of the roll-back length and 
a sudden deterioration of the run-time behavior when the number 
of computing processors is increased.

There are several aspects of PDES algorithms that should be 
considered in systematic efficiency studies. Some important 
aspects are: the synchronization procedures, the utilization of 
the parallel environment as measured by the fraction of working 
processors, memory requirements, inter-processor communications 
handling, scalability as measured by evaluating the 
performance when the number of computing processors becomes large, 
and the speed-up as measured by comparing the performance with 
sequential DES. In routinely performed studies to date, the efficiency 
is investigated in a heuristic fashion by testing the performance of a 
selected application in a chosen PDES environment (i.e., in a parallel 
simulator). Recently, Korniss {\it et al}. \cite{KTNR00} introduced a novel 
and powerful approach in which a PDES algorithm can be studied in an 
abstract way by extracting key features of the algorithm, 
simulating its performance, and applying the methods of 
non-equilibrium surface growth \cite{BS95} to evaluate 
its theoretical efficiency. In the Korniss {\it et al}. approach, 
the main concept is the simulated time horizon (STH), 
defined as the collection of LVTs of all logical processes. 
The growth rule of this virtual time surface is defined 
by the communication rule among logical processes (i.e., by 
their communication topology, which in turn is defined by the 
underlying dynamics of the physical system being simulated) 
and by the way in which the algorithm handles the advances in LVTs. 
In this picture, the utilization of the parallel environment 
is evaluated as the mean density of local update sites of the growing time 
interface, and the width of the interface at saturation provides 
a measure of desynchronization that is directly related to 
the memory requirements \cite{KNKG02}. Scalability 
properties of a PDES algorithm can be assessed from these 
performance simulation studies \cite{KTNR00,KNKG02,KNK03}.

In the study of the STH generated by a conservative PDES \cite{KTNR00},  
it has been determined that in the worst-case 
conservative scenario for a closed spin chain, when each 
processing element (PE) carries only one spin site (i.e., 
each logical process simply corresponds to the flipping 
of one spin) and communicates only with its nearest neighbors, 
the time evolution of the STH on coarse-grained scales is 
governed by the Kardar-Parisi-Zhang (KPZ) stochastic equation 
\cite{KPZ86}. This proves, by universality arguments, that 
the simulation phase of conservative PDES is asymptotically 
scalable, which guarantees a nonzero utilization even for an 
infinite number of PEs. Using the same argument, it has been 
shown that the STH becomes infinitely rough in the limit of an 
infinite number of PEs, 
which suggests possible difficulties with data management.  
Thus, the measurement phase of conservative PDES is 
not asymptotically scalable \cite{KNKG02}. Recent simulation 
studies \cite{KNK03} show that conservative PDES can be made 
fully scalable when the algorithm is supplemented with either 
a moving time window constraint \cite{Lub88b,Nic91} or 
additional scale-free communication patterns between PEs \cite{KNG+03}.

From the physics point of view, the virtual time surface of the 
generic conservative PDES, with its morphology and dynamics, 
can be viewed as a surface growing through deposition of random time 
increments in accordance with a growth rule defined by a generic 
conservative PDES update rule. The physics of this growth is reflected 
in the utilization (the fraction of non-idling PEs) that corresponds to 
the mean number of deposition events on the surface. 
In the case of a closed spin chain this is equivalent to 
the mean density of local minima in the interface. It should be 
possible, at least for steady-state simulations, to make 
an explicit connection between the utilization and the microscopic 
structure of the interface. Such a connection would enable   
rigorous studies of the update statistics and  
a closed  theoretical formula for the utilization. The coarse-grained 
methods previously applied to this problem \cite{KTNR00} provide 
a proof of asymptotic 
scaling properties in the limit of a large number of PEs. 
Because of their continuum nature they cannot give 
a detailed microscopic description of the interface; 
neither is it certain if their results are valid for statistically 
feasible moderate to large numbers of PEs. On the other hand, 
the mean utilization strictly depends on the microscopic structure of the 
STH. In this paper we explore the connection 
between the STH interface morphology on the micro-scale 
and the update statistics by addressing the above questions. 
Recently, similar connections have been established between the interface 
microstructure and its mobility for Ising and solid-on-solid models 
with various dynamics \cite{Rikv00,Rikv02a,Rikv02b,Rikv03}.

Section~\ref{model} outlines the simulation algorithm for modeling the 
generic conservative PDES of spatially decomposable cellular 
automata when each PE carries $N$ lattice sites. The steady-state 
update statistics for $N=1$ is analyzed in Sec.~\ref{statistics}. 
Here we derive formulas for the theoretical utilization and 
the theoretical probability distribution of updates in the system 
within an approximate model.  
Our approach uses simple statistics to build and to count distinct 
surface configuration classes consistent with a model update rule 
(or deposition rule). The idea may be generally applied to any 
surface that grows on a lattice by a known growth rule. When the 
growth recipe is known, it is possible to construct diagrams of  
local lattice-site configurations and to translate the 
rule to dependences among the graphs. Then the event 
probability on the surface is deduced from the corresponding 
diagram of possible surface-configuration classes. 
The performance of conservative algorithms is discussed in 
Sec.~\ref{performance}, where the results of Sec.~\ref{statistics} 
are applied to estimate the theoretical computational speed-up for the ideal 
system of PEs in a ring communication topology.
In Sec.~\ref{discuss} we discuss 
generalizations of our approach to other growth processes and 
advantages that follow in terms of practical applications such 
as, e.g., the possibility of computing closed-form expressions for quantities that 
would be unavailable by standard approaches.

\section{MODEL SIMULATIONS OF CONSERVATIVE UPDATE EVENTS
\label{model}}

\begin{figure}[bp]
\includegraphics[width=8.0cm]{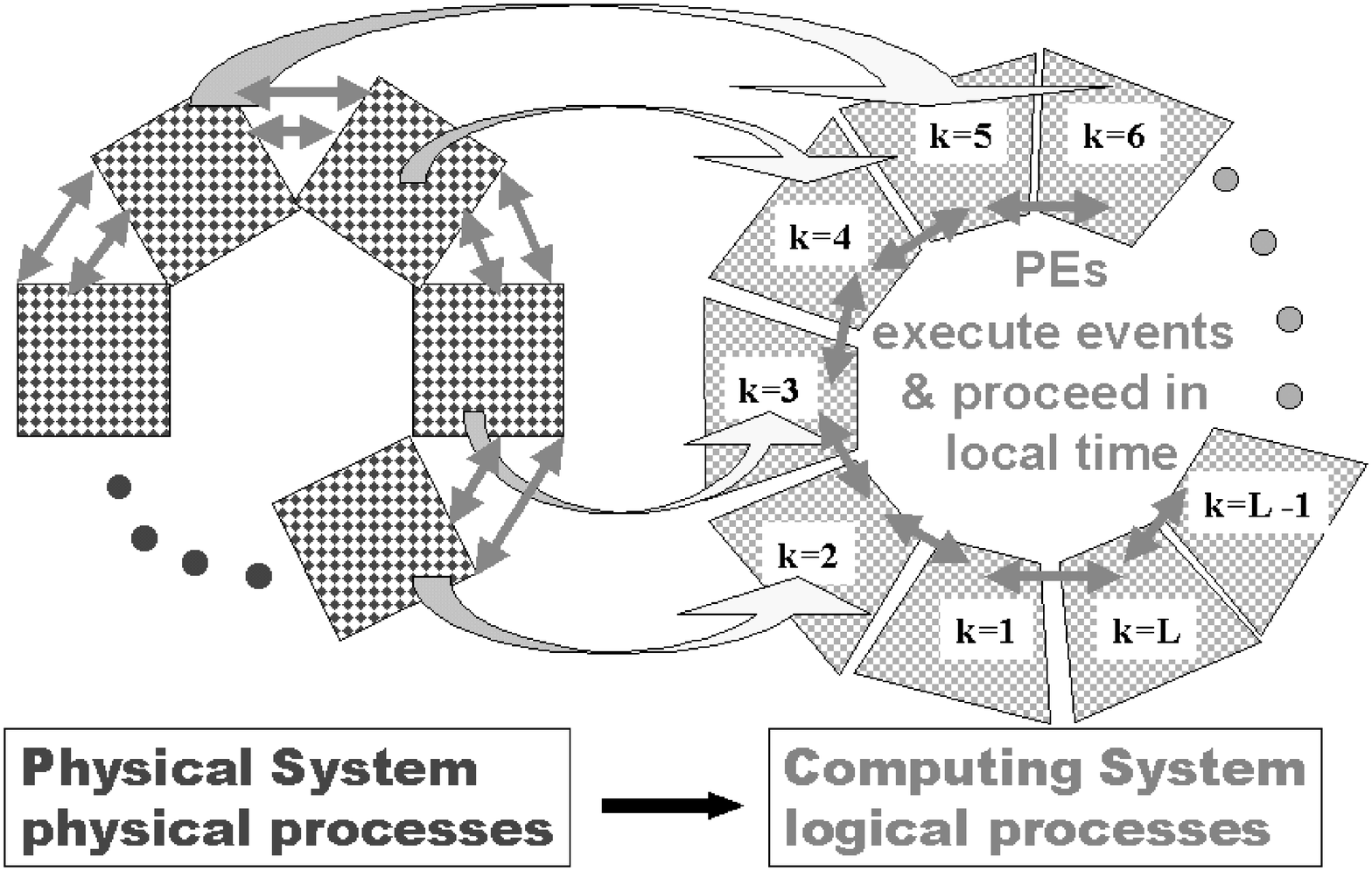}
\caption{\label{fig01} The mapping of physical processes to logical 
processes considered in this work. The nearest-neighbor physical 
interactions (two-sided arrows in the left part) on a lattice with 
periodic boundary conditions are mapped to the ring communication 
topology of logical processes (two-sided arrows in the right part). 
Each PE carries $N$ lattice sites, but communications take place 
only at border sites. In this study, each PE has at most two effective 
border sites.}
\end{figure}

We consider an ideal system of $L$ processors, arranged 
on a ring (Fig.~\ref{fig01}). 
As an ideal system we understand a system of identical PEs, 
where communications between PEs take place instantaneously.
Each PE carries $N$ lattice 
sites, $N_b$ of which are border sites and $(N-N_b)$ are 
interior sites (where all immediate lattice neighbors reside on 
the same PE). On each PE the simulation algorithm randomly selects 
one of the $N$ sites. If the selected site is a border site, 
the PE is required to communicate with its immediate neighbor(s)  
in an update attempt. If an interior site is selected, the 
update happens without communication between PEs. For this 
system, a discrete event means an update attempt. The state 
of the system does not change between update attempts. Processing 
elements perform operations concurrently. However, update attempts 
are not synchronized by a global clock. 

An example of the kind of system described above is a large, 
spatially extended ensemble of spins, arranged on a regular lattice, 
with a concurrent operation of random Monte Carlo spin-flip attempts. In this 
picture, the ensemble is spatially decomposed into $L$ subsystems, 
each of which carries $N$ spin sites. Each subsystem is placed on a PE, 
and the required communication is the exchange of information 
about states of the border spins (Fig.~\ref{fig01}). 
In the simplest case of $N=1$, 
the system is a closed spin chain, and the spin-flip attempt at 
the $k$th PE depends on the two nearest-neighbor spins located on 
the $(k-1)$th and the $(k+1)$th PEs. The $k$th PE is not allowed 
to update until it receives information from the neighboring PEs. 
For general $N$, a sublattice assigned to a PE has $N_b$ border 
spins. However, for example in Monte Carlo simulations, at each update 
attempt only one of the border sites may be randomly selected at 
a time: either a site from the left border slice or a site from 
the right border slice. Therefore, considering communications between 
logical processes, there are only two effective border sites per  
PE when $N \ge 2$. The case when $N>1$ and the effective $N_b=1$, 
is realized when on each PE the left and the right border slices coincide. 
This case is equivalent to a closed spin chain, i.e., to the case of $N=1$.

In generic conservative PDES, to simulate asynchronous dynamics 
employing $L$ processors, the $k$th PE generates its own local 
simulated time $\tau_k$ for the next update attempt. 
The $k$th local simulated time models the LVT of the $k$th logical process. 
Update attempts 
are simulated as independent Poisson-random processes, in which 
the $k$th random time increment $\eta_k$ (i.e., the random time 
interval between two successive attempts) is exponentially 
distributed with unit mean. A processor is allowed to update its 
local time only if the update is guaranteed not to violate 
causality. Otherwise, it remains idle. The time step $t$ is the 
index of the simultaneously performed update attempt. 
It corresponds to an integer wall-clock time with each PE 
attempting an update at each value of $t$. Explicitly, 
in our model simulations the 
generic conservative update rule allows the $k$th PE to update 
at any time step $(t+1)$ if either of two conditions is satisfied. 
One: the randomly chosen lattice site is in the interior. 
Two: the randomly chosen lattice site is a border site and either 
of the following update conditions is satisfied:
\begin{eqnarray}
\label{rule1}
N=1 & : & 
\tau_k (t) \le \min \left\{ \tau_{k-1} (t), \tau_{k+1} (t) \right\}, \\
\label{rule2}
N \ge 2 & : & 
\tau_k (t) \le \tau_r (t),
\end{eqnarray}
where $r=k-1$ when the left border site is chosen and $r=k+1$ when 
the right border site is chosen. Following a successful update attempt, the 
local simulated time is incremented for the next update attempt: 
$\tau_k(t+1)=\tau_k(t)+\eta_k(t)$. The random time increment 
$\eta_k(t)$ is computed at each $k$ and $t$ as $\eta_k(t)=-\ln(r)$, 
where $r \in (0;1]$ is a uniform deviate. The periodicity condition 
requires communication between the first and the last PEs in the 
chain: $\tau_{L+1}(t)=\tau_1(t)$. In simulations we iterate either the 
update rule (\ref{rule1}) or the update rule (\ref{rule2}), 
starting with the initial condition 
$\tau_k(t=0)=0$ for all $k$.

\begin{figure}[tp]
\includegraphics[width=6.5cm]{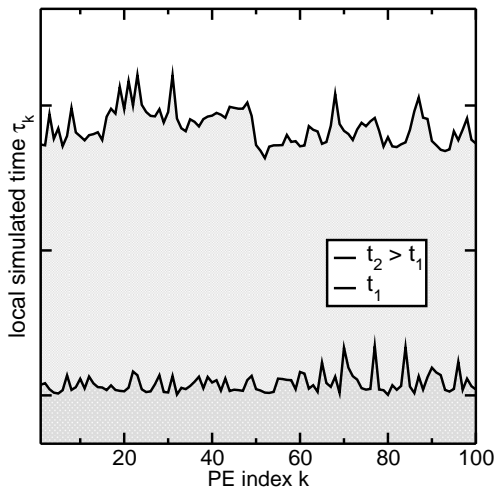}
\caption{\label{fig02} The growth and roughening of the STH for 
$L=100$ and $N=1$: snapshots at $t_1$ (lower surface) 
and $t_2$ (upper surface). 
Here, $t_1 < t_2 \ll t_{\times} \approx 3700$. Local heights 
$\tau _k$ are in arbitrary units.}
\end{figure}

For the set of $L$ processing elements, we define the STH as the set 
of $L$ local simulated times at time step $t$. The mean height of 
the STH is given by the mean virtual time 
$\langle \tau (t) \rangle _L = 1/L \sum\nolimits^L_{k=1} \tau_k(t)$. 
Figure \ref{fig02} presents the STH generated for a closed chain 
of $L=100$ processors. As the time index advances the STH 
grows and roughens. The time evolution of the statistical spread 
of the interface is characterized by two distinct  
phases, the growth phase (when $t \ll t_{\times}$) and the saturation phase 
(when $t \gg t_{\times}$), separated by the cross-over time $t_{\times}$. 
For a finite $L$, $t_{\times}$ marks the transition to the steady state, 
where the average width of the interface is constant in time and is given by 
the power law $(NL)^{1/2}$ \cite{KTNR00,KNK03b}. 

\begin{figure}[tp!]
\includegraphics[width=6.5cm]{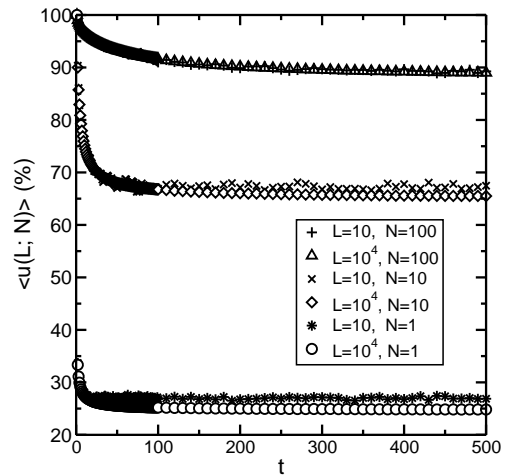}
\caption{\label{fig03} The time evolution of the utilization 
$\langle u(t) \rangle$ (averaged over $K=1024$ simulations) for 
$L=10$ and $10^4$ with $N=1, 10, 100$. The result depends most strongly on $N$.}
\end{figure}

To study the parallel efficiency, 
we define the utilization $u(t)$ as the fraction of PEs 
that perform an update at the parallel time step $t$. The simulated 
utilization $\langle u(t) \rangle$ is computed as an ensemble average 
over many independent simulations. The time evolution of the simulated 
utilization reaches a steady state $\langle u(t) \rangle = \mathrm{const.}$  
that depends on the system size (Fig.~\ref{fig03}): the steady-state utilization 
grows monotonically with $N$.  
Note, for $N=1$, according to the conservative update rule~(\ref{rule1}), 
at $t$ the update at the $k$th PE site 
does not happen unless its cumulative local simulated time 
after $(t-1)$ steps is not larger than the cumulative local 
simulated times at its neighboring PE sites. This means that an update 
at the $k$th PE site corresponds to a local minimum of the STH at 
the $k$th site. Accordingly, the mean utilization 
$\langle u(t) \rangle$ represents the mean number of local minima 
in the STH interface at $t$, averaged over many independent 
simulations. In an individual simulation, the 
utilization $u(t)$ is the density of the local minima in the 
STH that is generated in this simulation. 
When $N \ge 2$ the utilization $u(t)$ is the density of updating 
sites in the interface. It is important to 
distinguish between $u(t)$ and $\langle u(t) \rangle$ 
as $u(t)$ is characteristic of a particular class of the STH 
configurations while $\langle u(t) \rangle$ is the  average 
measurement of $u(t)$ taken over all possible configuration 
classes. In analyzing the steady-state update statistics the 
steady-state utilization is denoted by $u$. 

\section{STEADY-STATE UPDATE STATISTICS FOR $N=1$
\label{statistics}}

The STH of the generic PDES can be identified with 
a one-dimensional ({\it 1d}) interface growing on a ring 
with the deposition of random time increments $\eta_k$ in 
accordance with the deposition (update) rule given by Eq.~(\ref{rule1}). 
The physics of this growth is reflected in the utilization. 
Because the utilization is strictly related to the microscopic 
structure of the interface it is possible to make an explicit 
connection between the utilization and the morphology of the STH 
and to derive an analytical formula for the theoretical mean utilization 
as a function of the system size $L$. In this section we make this 
connection for $N=1$ when the STH growth has reached the saturation 
phase (i.e., when $t \gg t_{\times}$). 
Our derivation of the update distribution makes the following two 
simplifying assumptions. First, we neglect correlations between  
nearest-neighbor local slopes. These depend on the type of 
deposition, i.e., our derivation is not specific to the distribution 
from which the deposited $\eta_k$ are sampled. This simplification is 
reflected in the assumption of equal statistical weights assigned to 
the legs of binary transition graphs that represent possible choices 
of neighboring local sites. Second, we neglect temporal correlations 
among the groups of the surface configuration classes. Because 
of the above two simplifications, our theoretical result for the mean 
utilization is a mean-field like approximation to the mean utilization 
measured in simulations.

\subsection{Theoretical utilization
\label{utilization}}

\begin{figure}[tp]
\includegraphics[width=7.5cm]{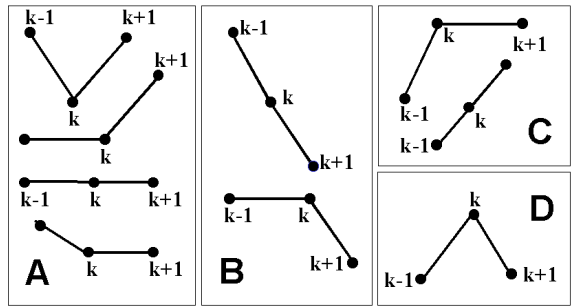}
\caption{\label{fig04} The four groups of elementary local surface 
configurations of the STH at the $k$th site. The index $k$ denotes 
the $k$th PE in the chain ($N=1$). Each group corresponds to one 
of the four mutually exclusive discrete events A, B, C and D at 
an update attempt. ``A" denotes an event when the update rule is 
satisfied. ``B" denotes an event when the update rule is not 
satisfied from the right. ``C" denotes an event when the update 
rule is not satisfied from the left. ``D" denotes an event when 
the update rule is not satisfied from the left and the right.}
\end{figure}

\begin{figure}[tp!]
\includegraphics[width=7.5cm]{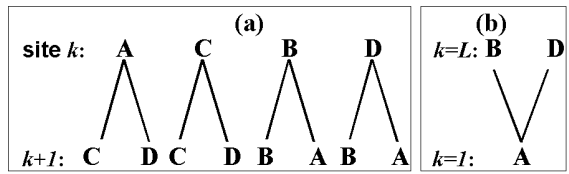}
\caption{\label{fig05} Binary branching of possible choices in 
constructing a surface configuration from the elementary local 
configurations A, B, C, and D of Fig. \ref{fig04}. (a) The 
alternatives that must be followed starting with A at $k=1$ and 
progressing towards $k=L$ to the right. (b) The only possible 
alternative for a periodic chain closed at $k=L$: the left 
neighbor of site $k=1$ must have configuration either B or D.}
\end{figure}

There are only four groups of elementary local site configurations of the STH 
that correspond to four mutually exclusive discrete  
events that take place at the $k$th PE site at $t$. 
These are as follows: ``A" denotes an event when the update 
rule~(\ref{rule1}) is satisfied from the left and from the right, 
i.e., when $\tau_{k-1} \ge \tau_k$ and $\tau_k \le \tau_{k+1}$; 
``B" denotes an event when the update rule~(\ref{rule1}) is 
not satisfied from the right, i.e., when 
$\tau_{k-1} \ge \tau_k$ and $\tau_k > \tau_{k+1}$; 
``C" denotes an event when the update rule~(\ref{rule1}) is not 
satisfied from the left, i.e., when 
$\tau_{k-1} < \tau_k$ and $\tau_k \le \tau_{k+1}$; 
and, ``D" denotes an event when the update rule~(\ref{rule1}) is 
not satisfied from either side, i.e., when 
$\tau_{k-1} < \tau_k$ and $\tau_k > \tau_{k+1}$. 
The corresponding elementary local configurations of the STH at the 
$k$th PE site are denoted by A, B, C and D (Fig.~\ref{fig04}). 
Because of the periodicity condition (i.e., $\tau_{L+1}=\tau_1$), 
during the steady state not all $L$ sites can have the same 
elementary site configuration \footnote{Strictly speaking, 
for $t>0$ the set of events when all sites are in the same 
elementary configuration A is of measure zero. In the absence 
of the periodicity condition, all sites could be in either 
B or C.}. 
Therefore, in the set of $L$ sites there must be at least one site 
with configuration A. Without losing generality, we assign the index $k=1$ 
to one of the sites that are in the local configuration A and enumerate the 
other sites accordingly, progressing to 
the right. Its right neighbor (having index $k=2$) can be only either 
in configuration C or in configuration D. Similarly, its left 
neighbor (having index $k=L$) can be only either in B or D. If 
site $k=2$ is in configuration C then site $k=3$  
can be only either in configuration C or D. If site $k=2$ is 
in D then site $k=3$ can be only either in B or A. 
These choices are presented as transition graphs (binary trees) 
in Fig.~\ref{fig05}. We adopt an approximation in which, during 
the steady state, the possible choices of transitions from the $k$th 
site to the right neighboring $(k+1)$ site are realized on average 
with equal frequency. 
Consequently, we assign equal statistical weights 
to each leg of the transition graph in Fig.~\ref{fig05}. 
Starting from the site $k=1$ 
and progressing to the right towards $k=L$, with the help of 
elementary transition graphs we can construct all possible configuration 
equivalency classes of the entire surface generated by the deposition 
(update) rule~(\ref{rule1}). These can be categorized into groups 
(called $p$-groups) based on the number $p$ of the deposition 
(update) events at $t$, i.e., the number of local minima 
in the surface configuration (coded by ``A") at $t$. The utilization of the 
$p$-group is $u(p)=p/L$. The probability distribution $f(p;L)$ 
of the deposition (update) events is obtained as the quotient 
of the multiplicity $M(p)$ of the $p$-group configuration class 
and the total number $M$ of configuration classes \footnote{
For convenience we drop the parameter $t$ in the notation when the 
analysis concerns the steady state.}.

\begin{figure}[tp!]
\includegraphics[width=7.0cm]{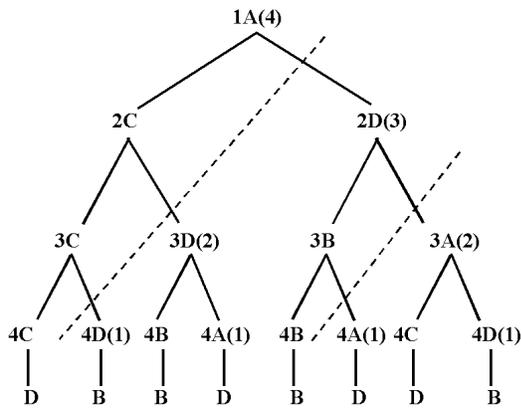}
\caption{\label{fig06} Binary tree for the construction of all 
possible configurations of the surface for $L=5$. A number to the left 
of the configuration symbol denotes the level of branching. A number 
in parenthesis to the right of the configuration symbol denotes the 
number of branching levels in a subtree. Notice the recurrent structure: 
the graph consists of the nested trees A(4), D(3), A(2), D(1). 
The dashed lines mark the transition cuts to the lower level trees. 
A(1) and D(1) denote the one level branch A and D, respectively, 
that mark the end of branching. See discussion in Sec.~\ref{utilization} 
and Appendix~\ref{derive}.}
\end{figure}

\begin{figure}[t!]
\includegraphics[width=7.0cm]{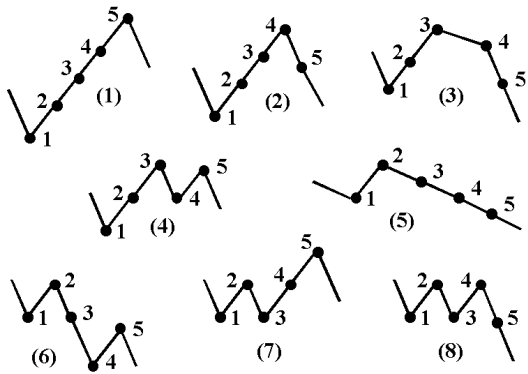}
\caption{\label{fig07} The graphs of possible surface configuration 
classes that correspond to the configurations read along 
the branches from Fig. \ref{fig06}: 1. ACCCD; 2. ACCDB; 3. ACDBB; 
4. ACDAD; 5. ADBBB; 6. ADBAD; 7. ADACD; 8. ADADB. 
Each graph represents a class of infinitely 
many topologically equivalent deformations.}
\end{figure}

For example, the binary tree for the construction of possible 
surface configuration classes for $L=5$ is shown in Fig.~\ref{fig06}. 
Looking along its branches, starting from the leading A at the fixed 
$k=1$ position, it is easy to identify a total of eight possible 
configuration classes of the entire surface: 1. ACCCD; 
2. ACCDB; 3. ACDBB; 4. ACDAD; 5. ADBBB; 6. ADBAD; 7. ADACD; 8. ADADB. 
Note, according to the surface construction rule, the class 
representative \#4 is not equivalent to the class 
representative \#7 because the leading A in configuration \#7 
has a local maximum as its right neighbor and configuration \#4 does 
not have this property. If the assignment of an index to a site 
were irrelevant, all configurations that can be obtained under an 
even permutation of sites would have fallen into one equivalency 
class. The surfaces representing configurations \#1-8 are sketched 
in Fig. \ref{fig07}. Each surface configuration represents a class 
of infinitely many topologically equivalent deformations because 
the deposited random time increment is a real positive number that  
can take on continuous values in the interval $[0; \infty)$. There are only two 
$p$-groups. In the first group there are four classes with one 
letter A: $M(1)=4$, $f(1;5)=1/2$ and $u(1)=1/5$. In the 
second group there are four classes with two letters A: 
$M(2)=4$, $f(2;5)=1/2$ and $u(2)=2/5$. Thus, for $L=5$ the mean 
utilization that is measured during steady-state simulations 
is $\langle u(L=5;N=1) \rangle = f(1;5) u(1) + f(2;5) u(2) =3/10$.

\begin{figure}[tp]
\includegraphics[width=7.5cm]{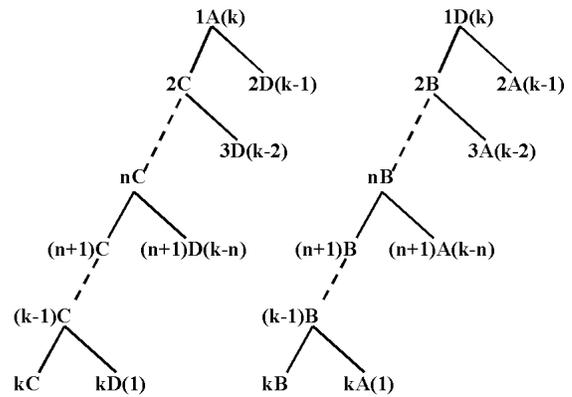}
\caption{\label{fig08} The recurrent structure of the binary tree 
in constructing the classes of surface configurations for  
general $L$. The meaning of the symbols is the same as in Fig.~\ref{fig06}. 
For general $L$, the highest level tree is 
$1$A$(L-1)$ that has $2^{L-2}$ branches. Each branch represents a 
class of surface configurations. The branches are categorized in 
distinct groups. Each group contains configurations with exactly 
$p$ repetitions of A$(1)$. The smallest $p$ is $1$, the largest 
$p$ is $\left[ \frac{L}{2} \right]$. The utilization in each 
group is $p/L$.}
\end{figure}

\begin{figure}[tp!]
\includegraphics[width=7.0cm]{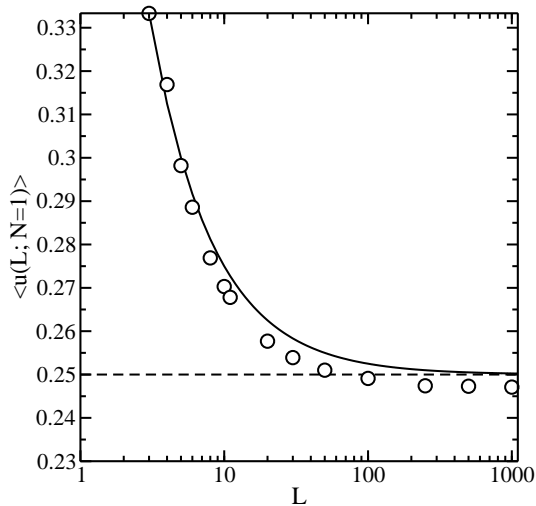}
\caption{\label{fig09} The steady-state mean utilization as a
function of the system size for $N=1$. The continuous curve represents
the analytical result (Eq.~(\ref{eq03})). It converges to
$\lim_{L \to \infty} \langle u(L;1) \rangle =1/4$ (horizontal line).
The circles represent the utilization measured in simulations, 
with error bars smaller than the symbol size.}
\end{figure}

For general $L$, the utilization measured in simulations during 
the steady state is the mean frequency of the local surface minima, 
averaged over all admissible surface configurations. It can 
be obtained from the generally valid formula for the computation 
of averages:
\begin{equation}
\label{eq02}
\langle u(L; N) \rangle = \sum_p f(p;L) u(p),
\end{equation}
where the summation extends over all $p$-groups of the admissible 
surface configuration classes, $u(p)$ is the utilization 
characteristic for each group, and $f(p;L)$ is the frequency 
of the occurrence of $p$-group during the steady state. 
To find the theoretical $f(p;L)$, one can exploit the recurrent 
structure of the corresponding binary tree (Fig.~\ref{fig08}) 
in counting the classes of the surface configurations (branches) 
that contain the elementary site configuration A at exactly 
$p$ number of sites, 
$p=1,2,3, \ldots ,p_{max} = \left[ \frac{L}{2} \right]$ 
($\left[ \frac{L}{2} \right]$ denotes the integral part, 
which is $L/2$ for even $L$ and $(L-1)/2$ for odd $L$). 
The details of the derivation are given in Appendix~\ref{derive}. 
The total number of configuration classes is $M=2^{L-2}$. 
The number of branches with exactly $p$ occurrences of A is 
$M(p)=(L-1)!/ \left( \left( 2p-1 \right) ! \left( L-2p \right) ! \right)$. 
The frequency of occurrence of the $p$-group is $f(p;L)=M(p)/M$. 
Thus, the theoretical mean utilization of the steady state is
\begin{eqnarray}
\label{eq03}
\langle u(L; 1) \rangle & = & \frac{1}{2^{L-2}}
\sum_{p=1}^{\left[ \frac{L}{2} \right]} 
{L-1 \choose 2p-1} \frac{p}{L} \nonumber\\
& = & \left\{ 
{ \begin{array}{cl}
1/2 & , L=2 \\ (L+1)/4L & , L \ge 3
\end{array} } \right. .
\end{eqnarray}
The theoretical utilization $\langle u(L;1) \rangle$ is bounded 
from below by $\langle u(L \to \infty ;1) \rangle = 1/4$ 
(Fig.~\ref{fig09}).

In classifying individual configurations, the underlying principle is provided  
by the deposition rule given by Eq.~(\ref{rule1}). Therefore, the local 
A-configuration represents four types of update events, and the local 
configuration B (or C) represents two types of no-update events 
(Fig.~\ref{fig04}). The small differences between the simulation results and 
Eq.~(\ref{eq03}), clearly observed in Fig.~\ref{fig09}, come mainly from 
neglecting temporal correlations among $p$-groups of surface configuration 
classes in our derivation. These correlations are intrinsically present in 
the computation of averages over time series in simulations but are absent 
in our model. They depend on the type of deposition, i.e., the probability 
distribution from which the random time increments $\eta_k$ are sampled. 
A possible second source of discrepancies is the assumption of equal 
statistical weights in the transition graphs (Fig.~\ref{fig05}). 
When the actual weights are only approximately equal, this modifies
the frequency $f(p;L)$ of the 
occurrence of a $p$-group in Eq.~(\ref{eq02}), so a particular 
surface configuration class may occur slightly more (or less) 
often in simulations than would result from our assumption. 
Note, this modifies only $f(p;L)$; the utilization $u(p)$ of a 
$p$-group is not changed. In deriving $f(p;L)$ the underlying 
assumption implies that any class of the entire surface 
configurations is equally probable. The factor $1/M=1/2^{L-2}$ in 
Eq.~(\ref{eq03}) has the meaning of this probability 
(Appendix~\ref{derive}).

\subsection{Computation of averages
\label{averages}}

\begin{figure}[!bp]
\includegraphics[width=8.0cm]{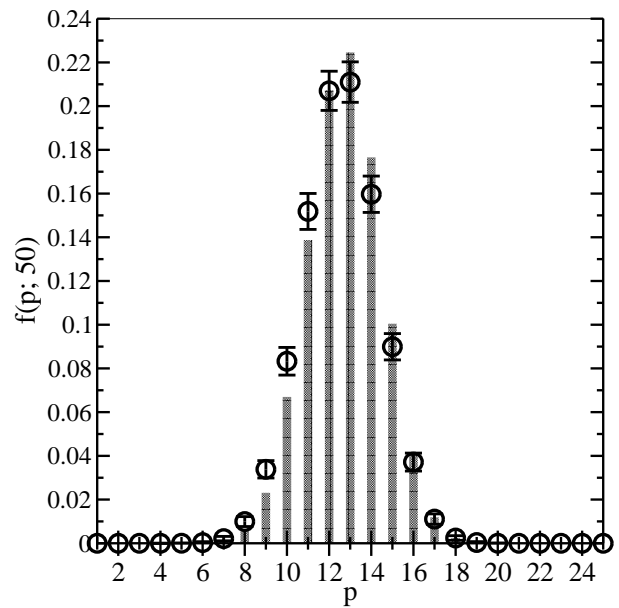}
\caption{\label{fig10} The probability distribution for $L=50$:
the theoretical $f(p;L)$ (histogram) and $\langle G(p;L) \rangle _T$ measured
in simulations (symbols). The error bars represent one standard
deviation from the mean of the measured time sequence at saturation 
(the quantity $\delta G(p)$ that enters Eq.~(\ref{eq04})).  
The measured frequencies were obtained from an ensemble
of $K=2048$ independent simulations as $K(p)/K$, where $K(p)$
is the number of trials that produced the $p$-group of the
surface configuration classes.}
\end{figure}

\begin{figure*}[!]
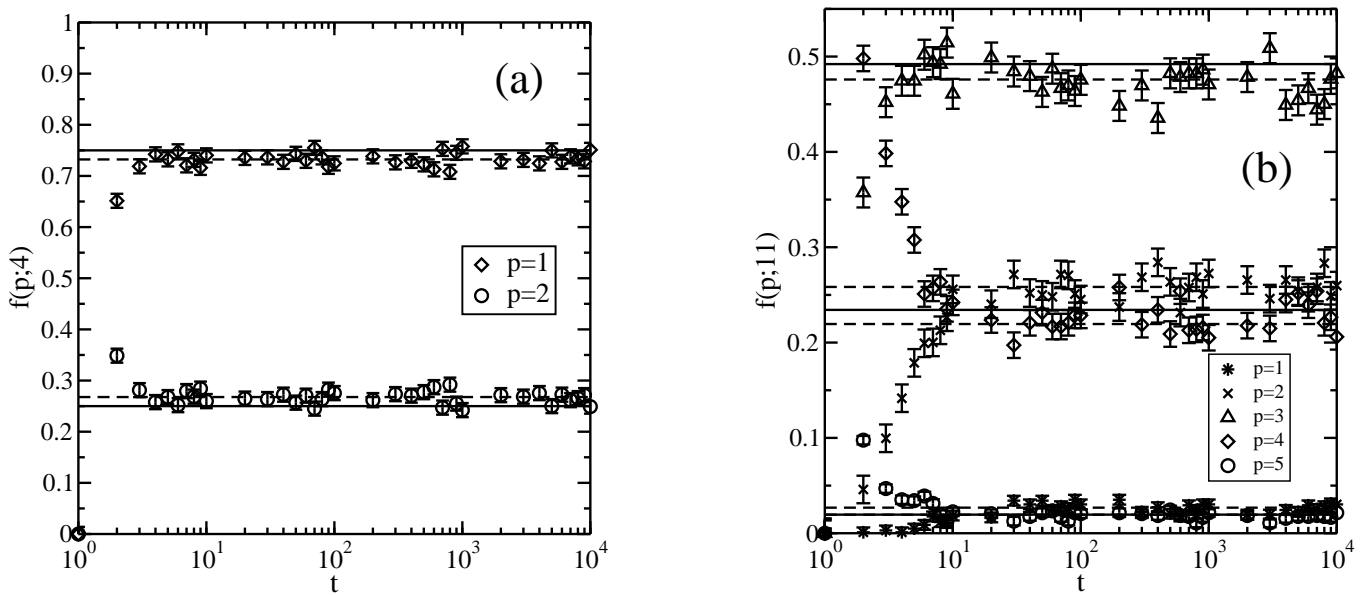

\includegraphics[width=8.0cm]{pre03-fig11a.eps}
\hfill
\includegraphics[width=8.0cm]{pre03-fig11b.eps}
\caption{\label{fig11} The time sequence of frequencies of the surface
configurations characterized by the utilization $u(p)=p/L$.
The continuous horizontal lines represent the theoretical $f(p;L)$.
Symbols are simulation data $G(p;L)$. The dashed horizontal
lines represent time averages $\langle G(p;L) \rangle _T$
over an interval of $1000$ $t$-steps, beginning at $t=10^8$.
The error bars represent one standard deviation from
$\langle G(p;L) \rangle _T$ as in Fig.~\ref{fig10}. The data were taken in $K=1024$
simulations: (a) For $L=4$: $\langle G(1;4) \rangle _T = 0.7323 \pm 0.0138$,
$\langle G(2;4) \rangle _T = 0.2677 \pm 0.0138$; (b) For $L=11$: 
$\langle G(1;11) \rangle _T =0.0268 \pm 0.0051$,
$\langle G(2;11) \rangle _T =0.2583 \pm 0.0144$,
$\langle G(3;11) \rangle _T =0.4759 \pm 0.0157$,
$\langle G(4;11) \rangle _T =0.2194 \pm 0.0134$ and
$\langle G(5;11) \rangle _T =0.0194 \pm 0.0044$.}
\end{figure*}

In simulations, the average steady-state utilization is measured at each $t$ 
as the arithmetic average over an ensemble 
of $K$ independent simulations and then averaged over a series of $T$ 
time steps during the steady state. At each $t$, this is equivalent to 
the computation of averages over the surface configuration classes in 
accordance with Eq.~(\ref{eq02}), where $f(p;L)$ is estimated from the 
steady-state simulation data. Denoting by $G(p;L)$ such 
an ``experimental" frequency, we write explicitly
\begin{equation}
\label{eq05}
\langle u(t) \rangle _K \approx \frac{1}{K} \sum_{i=1}^K u(i,t)
=\sum_{p=1}^{\left[ \frac{L}{2}\right] } G(p,t) u(p),
\end{equation}
where the rhs follows simply from grouping the summation terms. 
This is possible because $u(i,t)$ takes on only the values $u(p)$ 
that characterize the $p$-group of the surface configuration classes. 
Having a sequence of measured 
frequencies $G(p;L)$ over the steady-state time interval, the 
time average $\langle G(p;L) \rangle _T$ can be computed for 
each $p$. After time averaging, Eq.~(\ref{eq05}) gives 
\begin{equation}
\label{eq06}
\langle u \rangle _{K, T} \approx
\sum_{p=1}^{\left[ \frac{L}{2}\right] }
\langle G(p) \rangle _T u(p).
\end{equation}
The corresponding statistical spread of the measured average 
utilization $\delta \langle u \rangle$, i.e., the standard deviation 
of the mean $\langle u(L) \rangle$, can then be determined from the 
measured standard deviations of $G(p;L)$:
\begin{equation}
\label{eq04}
\delta \langle u(L) \rangle \approx \sqrt{\sum_p \left[ u(p) \delta G(p) \right]^2 },
\end{equation}
where $\delta G(p)$ denotes the empirical standard deviation of the 
$G(p;L)$ time sequence. At each $t$, the frequencies $G(p;L)$ are 
found by directly counting the simulations that produced 
$u(p)=p/L$ and, subsequently, computing the quotient of this 
count $K(p)$ and the total number $K$ of simulations in an ensemble. 
Explicitly, for $p=1,2,\ldots,\left[ \frac{L}{2} \right]$, 
the measured frequency is $G(p;L)=K(p)/K$, where 
$K=\sum_p K(p)$ (Fig.~\ref{fig10}).

A typical time sequence of $G(p;L)$, measured in $K=1024$ independent 
simulations, is shown in Fig.~\ref{fig11}. For $L=4$, the theoretical 
steady-state frequencies, $f(1;4)=3/4$ and $f(2;4)=1/4$, differ slightly from 
the averages $\langle G(p;L) \rangle _T \pm \delta G(p)$   
computed over an interval of $T=1000$ steps, beginning at $t=10^8$.  
The measured steady-state utilization 
$\langle u(L;1) \rangle \pm \delta \langle u(L) \rangle $ is  
$\langle u(4;1) \rangle =0.3169 \pm 0.0077$. 
Similarly, for $L=11$ the measured frequencies are in close agreement 
with the theory: 
$f(1;11)=f(5;11)=5/256$, $f(2;11)=f(4;11)=15/64$ and 
$f(3;11)=63/128$. 
The measured steady-state utilization is: 
$\langle u(11;1) \rangle =0.2678 \pm 0.0073$. 
The theoretical utilizations $\langle u(4;1) \rangle =5/16$ and 
$\langle u(11;1) \rangle=3/11$ (from Eq.~(\ref{eq03})) compare 
with the utilizations measured in simulations well within 
the statistical error bars when $K=1024, 2048$; likewise, there is very 
good agreement for general $L$. However, when $K=4096$ the statistical spread 
$\delta \langle u(L) \rangle$ is small enough 
to see that the results of Eq.~(\ref{eq03}) lie above the simulation data 
in Fig.~\ref{fig09}.

The standard deviation of the distribution of $u(p)$ among 
admissible $p$-groups of the surface configuration classes can be 
measured directly in simulations as the square root of the variance $\mathrm{var}(u)$: 
\begin{eqnarray}
\label{eq07}
\mathrm{var}(u) & \approx & 
\langle u^2 \rangle _{K, T} - \langle u \rangle _{K, T}^2 \\
& = & \sum_{p=1}^{\left[ \frac{L}{2} \right]} \langle G(p;L) \rangle _T 
u(p)^2 - \left(
\sum_{p=1}^{\left[ \frac{L}{2} \right]} \langle G(p;L) \rangle _T u(p)
\right) ^2,\nonumber
\end{eqnarray}
where the rhs comes from grouping terms in the summations. 
The statistical spread in $u(p)$ is the largest for $L=4$ and decreases 
when $L$ increases. 
Equation~(\ref{eq07}) gives the measured average variance of the 
probability distribution of updates in the system, scaled by $L^2$.

\subsection{Probability distribution of updates
\label{distribution}}

\begin{figure}[tp!]
\includegraphics[width=8.0cm]{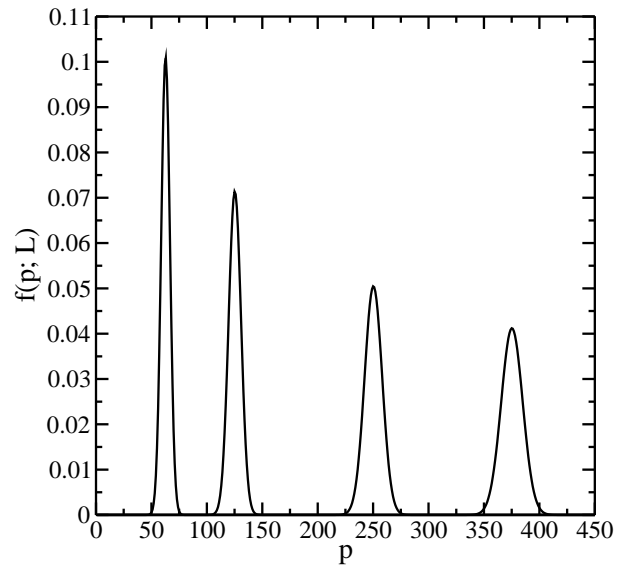}
\caption{\label{fig12} The probability distribution 
$f(p;L)$ of $p$ updates in a closed linear chain of 
$L$ PEs, each carrying one lattice site and following the 
conservative update rule. $L=250,500,1000,1500$ (from left 
to right).}
\end{figure}

The derived theoretical probability distribution of updates in 
a closed linear chain of $L$ processors, each carrying $N=1$ lattice 
sites and following the conservative update rule, is
\begin{equation}
\label{eq08}
f(p;L) = \frac{1}{2^{L-2}} {L-1 \choose 2p-1},
\end{equation}
where $p$ is the number of updates at the $t$th update attempt 
in the steady-state simulation. In an equivalent interpretation, 
Eq.~(\ref{eq08}) gives the probability of generating a surface 
with $p$ local minima at saturation when the surface 
growth obeys rule~(\ref{rule1}). In other words, in the latter 
interpretation $f(p;L)$ is the probability distribution of the 
deposition events on the surface. 
Equation~(\ref{eq08}), derived in Appendix~\ref{derive}, was already 
used in calculating $\langle u(L;1) \rangle$ in Eq.~(\ref{eq03}). 
Figure~\ref{fig12} presents 
$f(p;L)$ for various system sizes. It can be seen that the measured 
utilization is $\langle u \rangle = \langle p \rangle /L$, where 
$\langle p \rangle$ is the mean of $f(p;L)$. The variance 
$\sigma^2$ of $f(p;L)$ (and, thus the standard deviation of $u$) 
can be obtained in the usual way (Appendix~\ref{moments}) as
\begin{equation}
\label{eq09}
\sigma^2 = \sum_m \left( m - \langle p \rangle \right)^2 f(m;L) 
=\frac{L-1}{16}
\end{equation}
for $L \ge 4$, and $\sigma^2=0$ for $L=2,3$; where
\begin{equation}
\label{eq10}
\langle p^2 \rangle = \sum_m m^2 f(m;L) 
=\frac{L(L+3)}{16}
\end{equation}
for $L \ge 4$, and $\langle p^2 \rangle=1$ for $L=2,3$; and
\begin{equation}
\label{eq11}
\langle p \rangle = \sum_m m f(m;L) 
=\frac{L+1}{4}
\end{equation}
for $L \ge 3$, and $\langle p \rangle=1$ for $L=2$. Also, 
using Eq.~(\ref{eq11}), it can be derived that $\langle 1/p \rangle=1$ 
for $L=2$, and for $L \ge 3$:
\begin{eqnarray}
\label{eq12}
\left\langle \frac{1}{p} \right\rangle & = & 
\sum_m \frac{1}{m} f(m;L) \nonumber\\
& = & \frac{8}{L(L+1)} \left( 
\sum_k \left(2k-1\right) f(k;L+2) 
- \frac{L+1}{2^L} \right) \nonumber\\
& = & \frac{4}{L} \left( 1 - \frac{1}{2^{L-1}} \right),
\end{eqnarray}
and for $L \ge 4$:
\begin{eqnarray}
\label{eq11a}
\langle p^3 \rangle &=&
\frac{L^3+6L^2+3L-2}{64},\\
\label{eq11b}
\langle p^4 \rangle &=& \frac{L(L^3+10L^2+15L-10)}{256}.
\end{eqnarray}
The form of Eq.~(\ref{eq08}) permits exact computations of
all moments of $f(p;L)$. The corresponding recursion formula
is given in Appendix~\ref{moments}. In this way we find that the skewness
of $f(p;L)$ is zero, which means that $f(p;L)$ is symmetric about
$\langle p \rangle$. The computed kurtosis (the fourth central moment) is
strictly positive for $L \ge 4$, which means that $f(p;L)$ is
more pointed than a Gaussian.

The theoretical standard deviation of the utilization distribution 
among various $p$-groups,  
i.e., the statistical spread of $u(p)$  
in an ensemble of independent simulations, can be computed from 
Eq.~(\ref{eq09}):
\begin{equation}
\label{eq13}
\sigma \left( u \right) = \frac{\sqrt{L-1}}{4L}.
\end{equation}
In an equivalent interpretation, Eq.~(\ref{eq13}) gives the distribution of 
$u(p)$ among various $p$-groups of surface configuration classes.

The statistical distribution of the updates in the system of 
$L$ processors can also be estimated directly from the simulation data, 
without any presupposed underlying model. 
It is sufficient to notice that the 
update at a PE site happens when the STH has a local minimum 
at that site and that the {\it 1d} surface of $L$ sites may have no 
more than $\left[ \frac{L}{2} \right]$ local minima. 
Denoting by $K(p)$ the number of simulations that produced surfaces with exactly $p$ 
local minima in the sequence of $K$  
independent simulations ($p$ and $K(p)$ are directly counted 
in simulations), at any $t$ the ``experimental" distribution is $G(p;L)=K(p)/K$. 
Its time average in the steady state is $\langle G(p;L) \rangle _T$. 
The variance of $\langle G(p;L) \rangle _T$ can be obtained directly from the simulation 
data as described in Sec.~\ref{averages} 
(by the rhs of Eq.~(\ref{eq07}) multiplied by $L^2$). In the infinite $K$-limit 
$\langle G(p;L) \rangle _T$ converges to the exact steady-state 
distribution $g(p;L)$. At any $t$ the average of 
any observable $Q$ that depends on the number of local minima can 
be evaluated as
\begin{eqnarray}
\label{eq14}
\langle Q \rangle _K & \approx & \frac{1}{K} \sum_{r=1}^K Q(r) \nonumber\\
& = & \frac{1}{K} \sum_{p=1}^{\left[ \frac{L}{2} \right]} K(p)
\left( \frac{1}{K(p)} \sum_{r=1}^{K(p)} Q(r) \right) \nonumber\\
& = & \sum_{p=1}^{\left[ \frac{L}{2} \right]} G(p;L) 
\langle Q \rangle _{K(p)},
\end{eqnarray}
where $\langle \cdot \rangle _{K(p)}$ is the mean over the 
measured configuration classes in the $p$-group. The exact mean 
in the steady state is 
\begin{equation}
\label{eq15}
\langle Q \rangle = 
\sum_{p=1}^{\left[ \frac{L}{2}\right] } g(p;L) Q(p,L),
\end{equation}
where $Q(p,L)$ is the value typical for the $p$-group.

The question that naturally arises at this point is how close the 
theoretical distribution $f(p;L)$ given by Eq.~(\ref{eq08}) represents 
the exact distribution $g(p;L)$ that enters Eq.~(\ref{eq15}). 
The results for $L=50$ obtained with $K=2048$, presented as 
a histogram in Fig.~\ref{fig10}, show that $f(p;L)$ mimics the 
overall shape of the experimental $\langle G(p;L) \rangle _T$ very well. 
Increasing $K$ to $4096$ improves the error bars of $\langle G(p;L) \rangle _T$  
(in particular, for the extreme values of $p$); yet, as $K$ gets 
larger the overall shape of $\langle G(p;L) \rangle _T$ remains unchainged and 
the difference $\langle G(p;L) \rangle _T-f(p;L)$ does not entirely vanish. 
This difference is most pronounced when $L$ is small and is  
largest for $L=4$. Since for $L=4$ there are only two values of 
$p$, it is easy to estimate the exact $g(p;L)$ to a high degree 
of confidence. For $K=1024$: $\langle G(1;4) \rangle _T=0.73226 \pm 0.01378$, 
$\langle G(2;4) \rangle _T=0.26773 \pm 0.01378$. 
For $K=5120$: $\langle G(1;4) \rangle _T=0.73186 \pm 0.00611$,  
$\langle G(2;4) \rangle _T=0.26813 \pm 0.00611$. As $K$ increases 
$\langle G(p;L) \rangle _T$ converges to: $g(1;4) \approx 0.732$,  
$g(2;4) \approx 0.268$; while the theoretical estimate is: 
$f(1;4)=0.75$, $f(2;4)=0.25$. The small difference $g(p;L)-f(p;L)$ 
for the worst case of $L=4$ indicates that Eq.~(\ref{eq08}) 
is a close approximation to the time-averaged exact distribution $g(p;L)$. 
We believe, the primary reason for this difference is the lack of temporal 
correlations among various $p$-groups in our computational model, 
as was pointed out in Sec.~\ref{utilization}. Regardless of 
this simplification, the theoretical standard deviations $\sigma (u)$ 
given by Eq.~(\ref{eq13}) agree with $\sigma (u)$ computed directly 
in simulations via Eq.~(\ref{eq07}).

\section{PERFORMANCE OF A CONSERVATIVE ALGORITHM
\label{performance}}

The computational speed-up $s$ of a parallel algorithm is 
defined as the ratio of the time required to perform 
a computation in serial processing on one PE to the time  
the same computation takes in parallel processing on $L$ processors. 
It is easy to derive from the above definition that for an ideal 
system of processors the computational speed-up is the product of 
the number $L$ of PEs in the system and the utilization 
$\langle u(L;N) \rangle$ of the parallel processing environment: 
$s=L \langle u(L;N) \rangle$. In other words, the speed-up is 
measured by the average number of PEs that work concurrently 
between two successive update attempts.

We observe that for ideal PEs the speed-up as a function $F(L)$ must be such that 
the equation $F(L)=s$ has a unique solution, where $s$ is a fixed 
positive number. This requirement follows naturally from  
the logical argument that distributing the computations over $L$ ideal 
processors gives a unique speed-up, i.e., two ideal systems having 
sizes $L_1$ and $L_2$, respectively, may not give the same $s$. 
This means that $F(L)$ must be a monotonically increasing function of $L$.

In our model for $N=1$ the average number of PEs that work in parallel, 
i.e., the speed-up, is the mean number of local minima in the 
STH during steady-state simulations. In Section~\ref{distribution} 
this number is computed explicitly as the first moment of a theoretical 
distribution given by Eq.~(\ref{eq08}). In this way, translating 
Eq.~(\ref{eq11}) to the language of applications, 
the theoretical speed-up that can be obtained in an ideal system 
of PEs performing conservative PDES in the ring communication 
topology with $N=1$, is given by:
\begin{equation}
\label{eq23}
s = \left\{
{ \begin{array}{cl}
1 & , L=2,3 \\ (L+1)/4 & , L \ge 4
\end{array} } \right. .
\end{equation}
In what follows we discuss the consequences of 
Eq.~(\ref{eq23}) (or its alternative Eq.~(\ref{eq03})) from the 
point of view of applications.

One of the consequences is that the theoretical 
upper bound for $\langle u(L;1) \rangle$ is $1/2$. This corresponds 
to the situation when only one of the two PEs is working at a time 
while the other one idles. In the picture when the simulations represent 
operations performed by a parallel algorithm, when $L=2$ or $L=3$ 
the parallelization within the conservative update scheme does not 
give an advantage in terms of the computation time because the 
processors work alternately. For an ideal system of PEs, 
where communications between PEs take place instantaneously, 
such an operation will not produce speed-up. 
For a real system of PEs, the communication overhead will 
produce an actual slow-down, i.e., the parallel execution time 
will be longer than the sequential execution time on one PE. 
Between the update attempts during the steady state the average number of 
PEs working in parallel is $L \langle u(L;1) \rangle=(L+1)/4$. 
This means, when $L=4$ or $5$ the actual number of working PEs is 
still either $1$ or $2$; and when $L=6$ or $7$ the actual number of 
working PEs is $1,2$, or $3$ at a time and on average this 
actual number is still either $1$ or $2$. This will produce a 
small speed-up for an ideal system of PEs, but for a real system 
of PEs this speed-up may be negligible or not present at all. 
A noticeable advantage in terms of speed-up should be expected 
when the average number of working PEs is $(L+1)/4 > 2$, 
which gives $L \ge 8$. For a real system of PEs the best 
speed-up will not be larger than the average speed-up for an 
ideal system, i.e., not larger than $s$ given by Eq.~(\ref{eq23}).

The parallel utilization efficiency $\langle u(L;N) \rangle$ and 
the speed-up, as measured by the average number 
$L \langle u(L;N) \rangle$ of PEs working in parallel, depend 
on the number $N$ of lattice sites per PE, as well as on the 
number $N_b$ of border lattice sites per PE, and on the 
communication topology among the PEs. Our earlier large-scale 
simulations \cite{KNK03} show that the worst-case 
scenario of $N=1$, studied in this work, can be greatly improved 
when $N$ is increased while retaining the ring communication 
topology with the effective $N_b=2$ (Fig.~\ref{fig03}). 
On the other hand, the case of $N_b=2$ seems to have limited 
applications and should rather be considered as an intermediate 
model towards the study of more realistic conservative PDES where 
the effective $N_b$ may be arbitrary. From this perspective the case 
of $N_b=2$ is really the best case scenario since the utilization 
declines with increased number of communications between PEs, 
i.e., when the effective $N_b$ increases. In the ring communication 
topology, it can be expected that the actual speed-up 
in the ideal system of PEs should be larger than it is 
in the worst-case scenario with $N=1$, but smaller than it is in 
the best-case scenario with large $N$ and $N_b=2$. The actual speed-up 
in the real system of PEs should not be larger than a 
theoretical upper bound for the ideal system. Deriving a 
theoretical estimate for this upper bound requires first 
finding a closed expression for $\langle u(L;2) \rangle$, 
a problem that is still awaiting solution.

\section{DISCUSSION
\label{discuss}}

We showed in Sec.~\ref{statistics} that for steady-state simulations 
for a closed chain it is possible to explicitly correlate deposition 
statistics with the surface morphology on a microscopic scale. 
In our approach we used simple laws of statistics to build distinct 
configuration classes of the virtual time horizon. For one 
particular rule of surface growth, we constructed binary trees from 
which we could read the surface equivalency classes, serving our purpose 
of counting a particular type of configurations, relevant in deriving 
an approximate analytical expression for the measured utilization, i.e., 
the measured frequency of the time deposition. In the case solved 
in this work, the summation process was technically easy once 
the symmetries in the binary graphs became apparent (Fig.~\ref{fig08}).  
In principle, our method may be applied to any surface that 
grows on a lattice by a known growth rule, and it can be generalized 
to any measurable quantity $Q$. When the growth recipe is available 
it should be possible to construct diagrams of elementary 
site configurations and to translate the growth rule to dependences 
among the graphs. An observable $Q$ can then be computed as the 
average over all available groups of classes of the surface 
configurations: $Q=\sum_p f(p)q(p)$, where $f(p)$ has the meaning 
of the probability distribution and $q(p)$ is the value of $Q$ 
characteristic for each group. In the example given in this work, 
the statistical weights assigned to each leg of an elementary transition 
diagram (Fig.~\ref{fig05}) were taken to be equal. However, a different 
deposition rule may require a different choice of elementary diagrams 
and a different choice of statistical weights. The present 
application of the method to the {\it 1d} deposition problem on the STH 
surface is a promising example that could be generalized to a 
variety of other growth processes.

In a common approach one finds the universal properties of 
growing interfaces from a stochastic growth equation that is 
solved in a coarse-grained approximation at large scales. One powerful 
technique is the renormalization group approach 
\cite{KPZ86,FNS77,MHKZ89,CMP98}. The coarse-grained solution of 
the stochastic dynamics provides asymptotic scaling properties 
in the limit of large system sizes. Because of its continuum 
nature this method is not capable of giving a detailed microscopic 
description of interfaces such as, e.g., the  probability distribution 
of events on the growing surface. When applied to the model of 
conservative PDES studied in this paper, the continuum method does 
not give any estimate of the utilization of the parallel environment and 
the speed-up. Neither does it give the scaling behavior for the 
utilization in the limit of infinite system size. In earlier work, 
Korniss {\it et al}. \cite{KTNR00} used a coarse-grained method to 
determine that, in the steady state for $N=1$, the conservative STH is 
governed by the Edwards-Wilkinson Hamiltonian \cite{EW82}, which implies 
a non-zero utilization in the infinite $L$ limit, i.e., the asymptotic 
scalability of a generic conservative PDES. This finding explained the 
observed tendencies in the time evolution of the large-scale simulation 
data for the utilization, which clearly showed that the steady-state mean 
utilization settles down at a non-zero value, slightly lower than $1/4$. 
The simplified analysis of the microscopic structure of the conservative STH 
at saturation, presented in this work, enabled us to derive the analytical 
formula for the utilization, Eq.~(\ref{eq03}), in the approximation where 
correlations among various surface configuration classes are absent 
during the steady state. Equation~(\ref{eq03}) provides the explicit scaling 
relation for the utilization, which shows directly the asymptotic scalability 
of conservative PDES. The limiting value 
$\lim_{L \to \infty} \langle u(L;1) \rangle = 1/4$ coincides with the  
estimate in Ref.~\cite{KTNR00}. 
The actual simulated values for the mean utilization 
$\langle u(L;1) \rangle$ fall below the analytical curve (Fig.~\ref{fig09}). 
This small, but statistically significant, difference is evidence for 
small spatial and temporal correlations inherently present in the simulation 
data during the steady state. Theoretical treatment of these correlations, 
which would provide a correction to the theoretical distribution derived in 
this work, remains to be explored.

The closed form of the event distribution during 
steady-state simulations (Eq.~(\ref{eq08})) enables one to compute analytically the 
mean of any observable that depends on the number of local minima in the 
STH surface (Eq.~(\ref{eq15})). In this way we derived the explicit  
expressions for $\langle p^n \rangle$ (Eqs.~(\ref{eq09})-(\ref{eq11b})) 
that are valid for all values of $L$ within the adopted model. 
When $N=1$, the measured mean number of local minima 
$\langle p \rangle$ directly translates 
to the utilization and to the speed-up of conservative PDES. 
This approach has an advantage over any of the common continuum methods 
because it gives not only the exact scaling relations in the limit 
of large system sizes, but also enables one to compute analytic  
quantities valid for any system size. 
While the asymptotic properties of a PDES (such as, e.g., 
the scalability as $L$ increases) are of theoretical 
interest, the explicit formulas for the performance evaluation of 
a PDES for finite $L$ and $N$ are of practical value in algorithm design.

When the number of lattice sites per PE is $N \ge 3$,
each PE carries two kinds of sites, interior sites and
border sites. This generates two groups of mutually exclusive
update events: updates when an interior site is
randomly chosen and updates when a border site is
randomly chosen. These groups of events are mutually exclusive
because during the $t$th update attempt the drawing of a lattice
site is performed only once per PE.
Finding the mean fraction of PEs that made an update while a border site
was selected requires solving the
case of $N=2$. This case is different from the case of $N=1$
because the update rule changes. Now, at each $t$
we first randomly select a border site on the $k$th PE, and
when the neighboring PE of the selected site has its local
simulated time larger than $\tau _k$  the $k$th PE makes the
update. This ``one-sided" rule changes the deposition pattern
(the STH growth rule) because now the $k$th PE site does not
need to be in elementary configuration A to make an update.
A new rule implies a new definition of $p$-group configuration classes that
now is characterized by $p$ updates (for $N=1$ it is defined by
$p$ local minima).
Also, the way in which new frequencies $f(p;L)$ are
computed  must incorporate a random choice of
border sites. We leave the questions of deriving update 
distributions for $N \ge 2$ open for future investigations.

\section{SUMMARY
\label{summary}}

We simulated the performance of an ideal closed chain of PEs that 
work in parallel in an asynchronous manner, with updates following 
a generic conservative algorithm. The conservative update rule can be 
seen as the mechanism that determines the growth of the virtual time 
surface of a conservative process. The physics of this growth 
is reflected in the utilization and in the interface width.

We showed that it is possible to make an explicit connection between 
the steady-state utilization and the microscopic structure of 
the virtual time interface at saturation. We exploited this
connection to derive an analytical formula for the probability 
distribution of the update events in the system within an approximate model. Then, having 
the model probability distribution, we computed explicit expressions for 
the mean utilization and the computational speed-up as functions of the 
system size. Our result states that the speed-up for the ideal closed 
chain of PEs, each carrying one lattice site, grows linearly with the 
system size as $s=(L+1)/4$ for $L \ge 4$, and $s=1$ for $L=2,3$.

Finally, we observe that our approach could be applied to a variety 
of other growth processes. In this sense, the present {\it 1d} application 
to the update problem in conservative PDES is a promising example. 
The main advantage of the approach is that it enables one to compute 
analytically quantities that otherwise can be only estimated 
qualitatively in an asymptotic fashion by continuum methods.

\begin{acknowledgments}
The authors thank G. Korniss for discussions. P. A. R. appreciates the 
hospitality of the MSU Department of Physics and Astronomy, and the ERC 
Center for Computational Sciences. 
This work is supported by NSF grants DMR-0113049 and DMR-0120310; and   
by the Department of Physics and Astronomy at MSU, and 
the ERC Center for Computational Sciences at MSU. 
This research used resources of the National Energy Research 
Scientific Computing Center, which is supported by the Office 
of Science of the US Department of Energy under contract No. 
DE-AC03-76SF00098.
\end{acknowledgments}

\appendix
\section{Derivation of $f(p;L)$
\label{derive}}

To find the theoretical frequency $f(p;L)=M(p)/M$, one should 
compute the number $M(p)$ of surface configuration classes 
that contain the elementary configuration A (a local minimum of the 
STH) at exactly $p$ sites. Both $M(p)$ and $M$ 
(the total number of configuration classes) can be easily 
found by simply counting the branches of the binary trees presented 
in Fig.~\ref{fig08}.

For a given $L$, we start with the highest level tree $1$A$(L-1)$ 
(the left tree in Fig.~\ref{fig08}). This tree has $(L-1)$ branching 
levels. The $1$A$(L-1)$ tree branches to $2$C$(L-2)$ and $2$D$(L-2)$ 
(the right tree in Fig.~\ref{fig08}), which have $(L-2)$ branching levels, 
etc. The branching ends up at $(L-1)$A$(1)$, $(L-1)$B$(1)$, 
$(L-1)$C$(1)$ and $(L-1)$D$(1)$ that do not branch. Therefore 
the total number of branches is $M=(1/2) \prod_{n=1}^{L-1} 2 = 2^{L-2}$.

The factor $1/M$ is the probability that any class of the entire 
surface configuration appears in an individual simulation at $t$. 
Assuming that each leg of the binary branch in Fig.~\ref{fig08} 
carries the statistical weight of $1/2$, this probability is $1/2^n$, 
where $n=L-2$ is the highest binary branching level.

To find $M(p)$, notice (Fig.~\ref{fig08} and Fig.~\ref{fig06}) 
that in each A-tree there is at least one sub-branch that 
contains no A (along the C-branch) and in each D-tree there is 
exactly one sub-branch that contains no A (along the B-branch). 
Let $n(p; k$X$(r))$ denote the number of branches that contain exactly 
$p$ number of As in the sub-branch $k$X$(r)$. Here, ``X" stands either 
for ``A" or ``D". In this notation the ``exactly one branch with no 
A in the D tree" means:
\begin{equation}
\label{ap01}
n \left( 0; kD(r) \right)=1.
\end{equation}
The $k$th level A-tree has $n(1; k$A$(r))$ branches that contain 
exactly one A:
\begin{equation}
\label{ap02}
n \left( 1; k A (r) \right)=1 + \sum_{s=1}^{r-1}
n \left( 0; \left( k+r-s \right) D (s) \right).
\end{equation}
The meaning of Eq.~(\ref{ap02}) can be clarified by the example 
of the tree in Fig.~\ref{fig06}. The first branch-cut (the 
left dashed line) separates the C sub-branch from the remainder. 
The C sub-branch (to the left of the first cut) has only one 
(leading) A. Therefore, there is $1$ on the rhs of Eq.~(\ref{ap02}). 
The remainder (to the right of the first cut) is the sum of the 
three D-trees: $4$D$(1)$, $3$D$(2)$,and $2$D$(3)$. Therefore, 
the sum in Eq.~(\ref{ap02}) has three terms. Since the remainder 
has already one A in the leading position, the sum must contain 
only the sub-trees that contain $p=0$ number of As. 
In a similar fashion we obtain for the D tree:
\begin{equation}
\label{ap03}
n \left( 1; k D (r) \right)= \sum_{s=1}^{r-1}
n \left( 1; \left( k+r-s \right) A (s) \right),
\end{equation}
where the summation extends over all A-trees because the left 
sub-branch of the D tree contains no As 
(Fig.~\ref{fig06}, Fig.~\ref{fig08}).

To find $n(2; k$A$(r))$, one must first sum over all D-trees 
(to the right of the first cut in Fig.~\ref{fig06} or along the 
left graph in Fig.~\ref{fig08}) and then sum over all A-trees 
(to the right of the second cut in Fig.~\ref{fig06} or along the 
right graph in Fig.~\ref{fig08}):
\begin{eqnarray}
\label{ap04}
n \left( 2; k A (r) \right) & = & 
\sum_{s=1}^{r-1} 
n \left( 1; \left( k+r-s \right) D (s) \right) \\
& = & \sum_{s=1}^{r-1} \sum_{l=1}^{s-1} 
n \left( 1; \left( k+r-l \right) A (l) \right) \nonumber \\
& = & \sum_{s=1}^{r-1} \sum_{l=1}^{s-1} 
\left( 1+ \sum_{m=1}^{l-1} 
n \left( 0; \left( k+r-m \right) D (m) \right) \right).\nonumber
\end{eqnarray}
In the last step of Eq.~(\ref{ap04}) we used Eq.~(\ref{ap02}).

In summary, in counting the branches with exactly $p$ number of As 
we utilize the recurrent structure of the binary trees presented in 
Fig.~\ref{fig08}. We perform the branch cuts that mark the transition 
from a higher level tree to a lower level tree and iterate along the 
branch-cut lines. The iteration is completed when the summation term 
has a form given by Eq.~(\ref{ap01}). In this way, continuing from 
Eq.~(\ref{ap02}), we obtain:
\begin{eqnarray}
\label{ap05}
M(1) & =& n \left( 1; 1 A (L-1) \right) \nonumber \\
& = & 1 + \sum_{s=1}^{L-2} 
n \left( 0; \left( L-s \right) D (s) \right) \nonumber \\
& = & 1+ \sum_{s=1}^{L-2} 1 =L-1, 
\end{eqnarray}
where we used Eq.~(\ref{ap01}). Similarly, continuing from 
Eq.~(\ref{ap04}), after substituting Eq.~(\ref{ap01}), gives:
\begin{eqnarray}
\label{ap06}
M(2) & =& n \left( 2; 1 A (L-1) \right) \nonumber \\
& = & \sum_{s=1}^{L-2} \sum_{l=1}^{s-1}
\left( 1+ \sum_{m=1}^{l-1} 1 \right) \nonumber \\
& = & \sum_{k=1}^{L-3} \frac{k(k+1)}{2} =
\sum_{k=0}^{L-4} {k+2 \choose 2} \nonumber \\
& = & {L-1 \choose 3}.
\end{eqnarray}
For $p=3$ the iteration leads to:
\begin{eqnarray}
\label{ap07}
M(3) & = & n \left( 3; 1 A (L-1) \right) \nonumber \\
& = & \sum_{s=1}^{L-2} 
n \left( 2; \left( L-s \right) D(s) \right) \nonumber \\
& = & \sum_{s=1}^{L-2} \sum_{l=1}^{s-1} 
n \left( 2; \left( L-l \right) A(l) \right) \nonumber \\
& = & \sum_{s=1}^{L-2} \sum_{l=1}^{s-1} \sum_{r=1}^{l-1} \sum_{k=1}^{r-1} 
\left( 1+ \sum_{m=1}^{k-1} 1 \right) \nonumber \\
& = & \sum_{k=1}^{L-5} \frac{k(k+1)(k+2)(k+3)}{2 \cdot 3 \cdot 4} 
\nonumber \\
& = & \sum_{k=0}^{L-6} {k+4 \choose 4} =  {L-1 \choose 5}.
\end{eqnarray}
where we used Eq.~(\ref{ap04}) and Eq.~(\ref{ap01}). 
For general $p=1,2, \ldots ,\left[ \frac{L}{2} \right]$, we obtain:
\begin{eqnarray}
\label{ap08}
M(p) & = & n \left( p; 1 A (L-1) \right) \nonumber \\
& = & \sum_{k_1=1}^{L-2} \ldots \sum_{k_{2p-2}=1}^{k_{2p-3}-1} 
\left( 1+ \sum_{k=1}^{k_{2p-2}-1} 1 \right) \nonumber \\
& = & \sum_{k=0}^{L-2p} {k+2p-2 \choose 2p-2} =  {L-1 \choose 2p-1}.
\end{eqnarray}

\section{Moments of $f(p;L)$
\label{moments}}

For the purpose of the computation of moments we
introduce the following notation:
\begin{equation}
\label{ba01}
\langle p^n \rangle _L = \sum_{m=1}^{\left[ \frac{L}{2} \right]}
m^n f(m;L),
\end{equation}
where $n=0,1,2,\ldots$. By manipulating the factorials in
$f(m;L)$ and shifting the summation index on the
rhs of Eq.~(\ref{ba01}), it is straightforward to show that:
\begin{equation}
\label{ba02}
\langle p^n \rangle _L = \frac{8}{L(L+1)}
\sum_{k=1}^{\left[ \frac{L+2}{2} \right]}
\left( k - 1 \right)^{n+1} \left( 2k-1 \right) f(k;L+2).
\end{equation}
With the substitution $L \to L-2$, Eq.~(\ref{ba02}) becomes
a recursion relation for $\langle p^n \rangle _L$:
\begin{equation}
\label{ba03}
\langle p^n \rangle _{L-2} \frac{(L-2)(L-1)}{8} =
\left\langle
\left( p - 1 \right)^{n+1} \left( 2p-1 \right) \right\rangle _L.
\end{equation}
Using the binomial formula in Eq.~(\ref{ba03}), the recursion
relation can be explicitly written out as:
\begin{eqnarray}
\label{ba04}
\langle p^{n+2} \rangle _L & = &
\langle p^{n} \rangle _{L-2} \frac{(L-2)(L-1)}{4^2} +
\frac{\langle p^{n+1} \rangle _L}{2}  \\
&+& \sum_{i=0}^n {n+1 \choose i} (-1)^{n+i+1}
\left( \langle p^{i+1} \rangle _L -
\frac{\langle p^i \rangle _L}{2}
\right).\nonumber
\end{eqnarray}
To obtain $\langle p^n \rangle _L $ for arbitrary $n$, 
we iterate Eq.~(\ref{ba04}), starting with the initial $n=0$ and
using the identities:
\begin{eqnarray}
\label{ba05}
\langle p^0 \rangle _L &=& 1,\\
\label{ba06}
\langle p^1 \rangle _L &=&\frac{L+1}{4}.
\end{eqnarray}
Equation~(\ref{ba05}) expresses the normalization condition. Equation 
~(\ref{ba06}) follows from Eq.~(\ref{ba01})  
after simple algebra and from Eq.~(\ref{ba05}).

For $n=0$, Eq.~(\ref{ba04}) gives:
\begin{equation}
\label{ba07}
\langle p^2 \rangle _L = \frac{L(L+3)}{4^2}.
\end{equation}
For $n=1$, Eq.~(\ref{ba04}) gives:
\begin{eqnarray}
\label{ba08}
\langle p^3 \rangle _L &=& \frac{(L-2)(L-1)}{4^2}
\langle p^1 \rangle _{L-2} \nonumber \\
&+& \frac{5 \langle p^2 \rangle _L}{2}
- 2 \langle p^1 \rangle _L - \frac{\langle p^0 \rangle _L}{2}.
\end{eqnarray}
Substituting Eqs.~(\ref{ba05})-(\ref{ba07}) leads to:
\begin{equation}
\label{ba09}
\langle p^3 \rangle _L = \frac{L^3+6L^2+3L-2}{4^3}.
\end{equation}
In a similar fashion, for $n=2$, Eqs.~(\ref{ba04})-(\ref{ba07})
and Eq.~(\ref{ba09}) give:
\begin{equation}
\label{ba10}
\langle p^4 \rangle_L=\frac{L\left(L^3+10L^2+15L-10\right)}{4^4}.
\end{equation}
The variance $\sigma^2$, the skewness, and the kurtosis of $f(k;L)$
can be computed in the standard way \cite{PTVF92}. The variance is 
given in Eq.~(\ref{eq09}). For the skewness we obtain $\mathrm{skew}(f)=0$. 
The kurtosis is a positive function of $L$. Explicitly, for $L \ge 4$:
\begin{equation}
\label{ba14}
\mathrm{kurt}(f)=2 \frac{3L^3+6L^2-10L+1}{(L-1)^2}.
\end{equation}

Equation~(\ref{eq12}) is derived in a similar way as
Eq.~(\ref{ba03}), by simple algebra and by shifting
the summation index. An arbitrary power $\langle p^{-n} \rangle$
can be obtained by deriving the corresponding recursion
relation, following the lines outlined above for $n>0$.

\end{document}